\documentclass[epjST]{svjour}
\usepackage{amsmath,amsfonts,graphicx}

\begin{document}
\title{Auto- and cross-correlations in the spinful topological Kondo model}
\author{Oleksiy Kashuba\inst{1}\fnmsep\thanks{\email{okashuba@physik.uni-wuerzburg.de}} }
\institute{$^1$Institut f\"ur Theoretische Physik und Astrophysik, Universit\"at W\"urzburg, 97074 W\"urzburg, Germany}
\abstract{%
The correlations in charge transport in multiple Majorana fermions systems contain much richer physics and carry more information than the conventional transport coefficients.
We calculate the auto- and cross-correlations in the spinful topological Kondo model and point out the frequencies dependencies of the correlations that may serve as a proof of the existence of the Majorana fermions in our system, particularly the change of the cross-correlation from positive to negative at higher frequency.
} 
\maketitle
\section{Introduction}
\label{sec:intro}

One of the major reasons for the constantly ascending interest to topological materials, such as topological insulators (TI) and topological superconductors (TSC), is the potential possibility to realize quantum computation~\cite{Nayak2008,Nilsson2008,FuKane2009,Sau2010} exploiting the braiding of anionic states, such as Majorana bound states (MBS)~\cite{Ivanov2001}.
In order to put this idea into effect, different methods of detection and manipulation of the Majorana states have to be thoroughly investigated.
The charge transport through these quantum states seems to be a natural way of accessing them.
The ordinary transport properties of a single Majorana state are well-studied~\cite{Mourik2012,Deng2016,Deacon2017a,Dominguez2017b,Pico2017a} especially in spin-orbit coupled quantum wires~\cite{Streda2003,Quay2010,Dominguez2012a,Heedt2017,Fleckenstein2018}, whilst the noise in many-Majorana systems still require a well-grounded investigation~\cite{Bolech2007}, since the cross-correlations contain much richer information about the internal entanglement of the quantum system and the statistics of excitations~\cite{Samuelsson2004,Heikkilae2013}.
The dependence of the cross-correlations between two Majoranas situated at the opposite ends of a wire on their coupling was already studied~\cite{Bolech2007,Nilsson2008,Lu2012}.
Following the Pauli principle one expects that the non-interacting fermions are antibunching, i.e. the noise cross-correlation are negative, what was also proven experimentally~\cite{Henny1999}.
Positive cross-correlations were predicted in case of the inelastic scattering~\cite{Rychkov2006} and demonstrated in the experiment~\cite{Oberholzer2006}.
The presence of interactions allows for positive cross-correlations as well~\cite{Samuelsson2002,Boerlin2002,Dominguez2010,Dominguez2012a}, what was later demonstrated~\cite{Neder2007}.

In this paper we investigate transport cross-correlations in the spinful topological Kondo model (STKM)~\cite{Kashuba2015}.
The motivation for our study was the recent experiment on the transport through a Majorana state with a mediating quantum dot~\cite{Deng2016}.
At certain parameters this system can be mapped onto the STKM (see section~\ref{sec:model} for details).
Measuring the spin currents (cross-) correlations one can justify the existence of the Majorana states at the end of the wire and extract additional information about the internal parameters of the system.


\section{Model}
\label{sec:model}

\begin{figure}
\centering
\includegraphics[width=.75\textwidth]{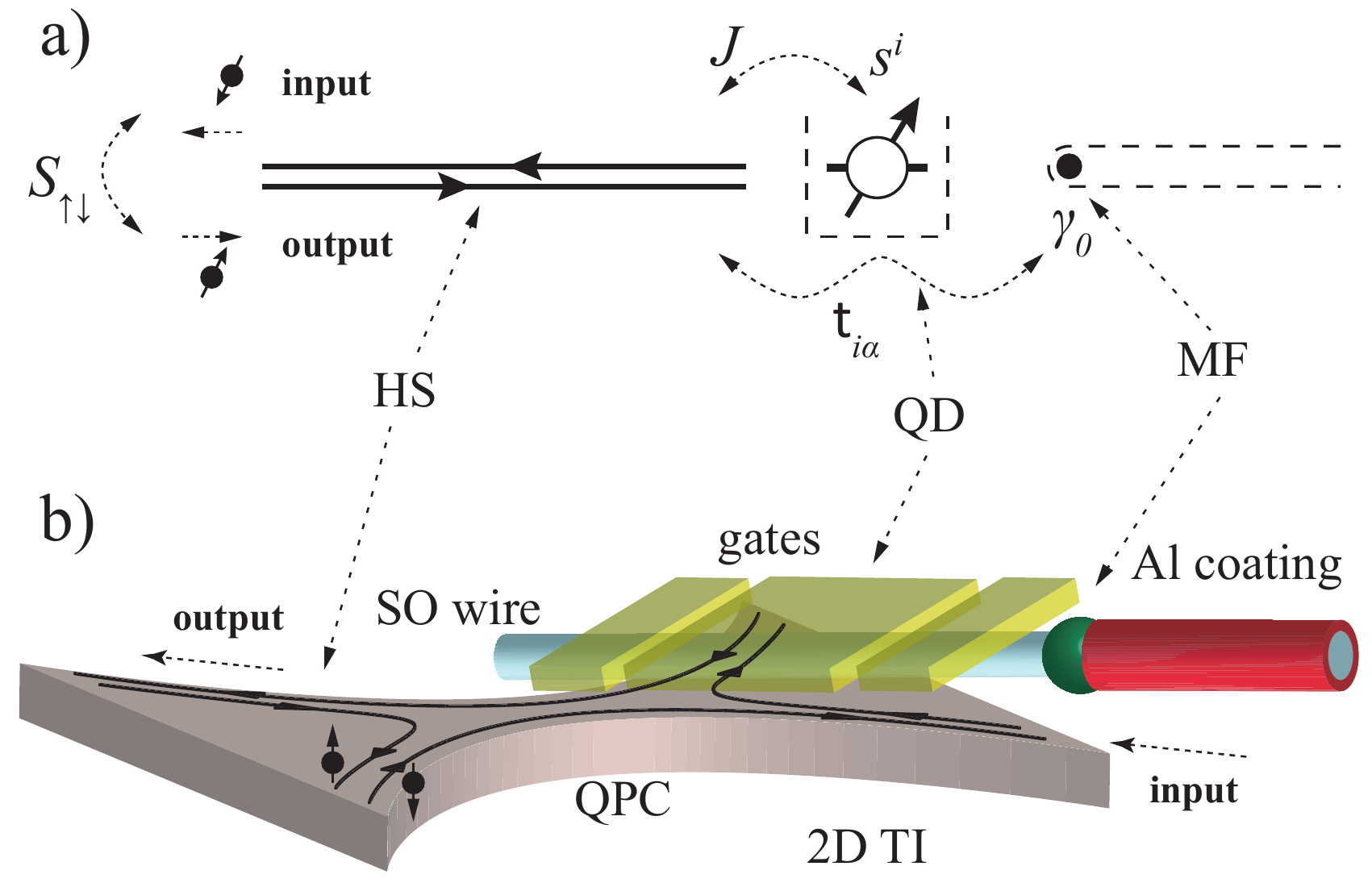}
\caption{The setup.
a) Schematical drawing of the setup. 
A Majorana fermion (MF) coupled to the Kondo quantum dot (QD), which in its turn is coupled to the propagating helical states (HS).
b) Physical realization of the setup.
The wire with the strong spin-orbit (SO) coupling is drawn by light blue color, dark red denotes an aluminum coating on the right side that induces the superconducting order in the wire.
A dark green sphere, where the coating ends, denotes a position of the Majorana bound state (MF).
The metallic gates used for the formation of the quantum dot in the SO wire under the middle gate~\cite{Deng2016} are drawn by semitransparent yellow color on the left of the Majorana state.
The 2D topological insulator with a quantum point contact (QPC) in a form of the thing bridge is drawn with the light brown color.
}
\label{fig:setup}
\end{figure}

The spinful topological Kondo model differs from the conventional Kondo by the additional tunnel coupling of the fermions to the Majorana states $\gamma_{i}$, in addition to the interaction between localized Majorana spin $s^{i}$ and electron spin density~\cite{Kashuba2015}.
The electrons Fermi sea is described by the Hamiltonian $H_{0}=\sum_{\alpha p}\xi_{p}c_{\alpha p}^{\dagger}c_{\alpha p}$ and is coupled to the Majorana fermions by interaction (biquadratic) and tunneling (bilinear) terms in the Hamiltonian
\begin{equation}
H= H_{0} + \frac{1}{2}\sum_{\alpha\beta,i} J_{i} s^{i}\, c_{\alpha}^{\dagger}\sigma_{\alpha\beta}^{i}c_{\beta}  +\sum_{i\alpha} \mathsf{t}_{i\alpha} \gamma_{i} c_{\alpha}^{\dagger} + h.c.
\end{equation}
Here, $c_{\alpha}=\sum_{p}c_{\alpha p}$ are electron ladder operators, $\xi_{p}$ is the energy of the Fermi sea excitations, indices $\alpha$ and $\beta$ run over the spin states $\uparrow,\downarrow$, index $i=1,2,3$, and the interaction coupled to the Majorana states is written in the most general form, since the Majorana spin components $s^{i}=-\frac i2 \sum_{jj'}\epsilon_{ijj'}\gamma_{j}\gamma_{j'}$ deplete all quadratic combinations of the Majorana ladder operators.
This Hamiltonian can be modeled in many ways, for example by the floating superconducting island, similar to the SO(N) TKM~\cite{Beri2012,Zazunov2013,Altland2014}, but the most experimentally relevant model is the combination of the quantum dot in the Kondo regime and a topological superconductor with a Majorana bound state~\cite{Lutchyn2014,Lee2013}.

The setup is schematically presented in Fig.~\ref{fig:setup} and based on the experimental setup from Ref.~\cite{Deng2016}.
The quantum dot is tunnel coupled to the wire with strong spin-orbit interaction and with the tunneling amplitude $\mathsf{t}_\text{TI--QD}$.
The quantum dot is coupled to the helical 1D states hosted by the 2D topological insulator (TI), instead of being directly coupled to the metallic lead~\cite{Posske2013,Posske2014}.
The thin bridge of the TI discriminates the left and right propagating modes, allowing, only one to tunnel though this quantum point contact (QPC).
This allows the separation the spin currents, since the excitations with the opposite spins propagate in the opposite directions and therefore should be detected in the different arms of the TI sample [see Fig.~\ref{fig:setup}b)].
If the quantum dot is in the Kondo regime, the strong Coulomb interaction $U_\text{QD}$ on the dot prohibits a simple tunneling to the dot, but allows a virtual exchange with the electrons in the helical wire, leading to the effective Kondo coupling $J\sim \mathsf{t}_\text{TI--QD}^{2}/U_\text{QD}$.
Thus, the quantum dot is occupied by the single electron and its dynamics can be fully described by the effective spin-1/2 operators $s^{i}$.
The Majorana state $\gamma_{0}$ tunnel coupled to the quantum dot with amplitude $\mathsf{t}_\text{QD--M}$ allows an additional virtual process---an electron hops on the dot, but leaves it tunneling into the Majorana states instead.
Such process effectively leads to the terms in the Hamiltonian $\mathsf{t}_{i\alpha} s^{i}\gamma_{0} c_{\alpha}^{\dagger} +h.c.$, where $\mathsf{t}_{i\alpha} \sim \mathsf{t}_\text{QD--M}\mathsf{t}_\text{TI--QD}/U_\text{QD}$.
The Fock space of the spin-1/2 and a Majorana state is equivalent to the Fock space of three Majoranas and their operators can be mapped onto the dot-Majorana setup as $\gamma_{i}=s^{i}\gamma_{0}$.

Since the spin conservation is broken, the only fixed point of the model is the strong coupling isotropic antiferromagnetic limit~\cite{Affleck1992,Hewson1997}.
One may show that for this type of spin-spin coupling only two out of six tunneling amplitudes $\mathsf{t}_{i\alpha}$ survive in the RG sense, leaving the tunnel coupling in the form $\sum_{i\alpha\beta} \mathsf{t}_{\beta} \sigma_{\alpha\beta}^{i} \gamma_{i} c_{\alpha}^{\dagger} + h.c.$, see Ref.~\cite{Kashuba2015}.

The tunneling coupling is not spin gauge invariant, and the Hamiltonian implicitly has a preferred axis of spin quantization.
This direction is defined by the details of the Majorana fermion construction, for instance, by the spin-orbit (SO) axis and the magnetic order/field $B$ direction in the nano-wire setup.
Choosing the spin gauge in which the spinor $\mathsf{t}_{\beta}$ takes the form $(\mathsf{t}/2,0)$, $\mathsf{t}\in\mathbb{R}$, we can write the Hamiltonian of our system in a form
\begin{multline}
H = H_{0}+\mathsf{t} \left( \frac12 \gamma (c_{\uparrow}-c_{\uparrow}^{\dagger}) - d c_{\downarrow}^{\dagger} + d^{\dagger} c_{\downarrow} \right)
+\\+ J\left(
c_{\uparrow}^{\dagger}c_{\downarrow} d^{\dagger}\gamma - c_{\downarrow}^{\dagger}c_{\uparrow} d\gamma \right)
+ J\left( c_{\uparrow}^{\dagger}c_{\uparrow} - c_{\downarrow}^{\dagger}c_{\downarrow} \right) \left(\frac12 - d^{\dagger}d\right)    
\label{eq:H}
\end{multline}
where $\gamma=\gamma_{z}$ and $d=(\gamma_{x}+i\gamma_{y})/2$.

The propagating 1D helical states can exist in the very same wire that is used for the constructing the quantum dot by the gates and the Majorana fermion by induced superconducting correlations~\cite{FuKane2009}, see Fig.~\ref{fig:setup}.
The spin gauge in the given Hamiltonian is same as the spin quantization axis of the helical states.
Such a matching allows to associate the left/right moving excitations with up/down spin channels in the Hamiltonian in Eq.~\eqref{eq:H}.
Thus, observing the incoming and outgoing electrons one may detect the spin channel correlations caused by the interaction effects in the topological Kondo model.

In the frame of the RG language, the tunneling terms are relevant, while the spin-spin interaction is marginal.
This difference motivates the assumption of taking into account $J$ perturbatively
and $\mathsf{t}$ exactly.
At $J=0$ the two subsystems, namely spin-up electrons $c_{\uparrow}$ plus Majorana fermion $\gamma$, and spin-down electrons $c_{\downarrow}$ plus conventional fermion $d$, are decoupled, leaving the spin components of the current uncorrelated.
To quantify the correlations, we calculate the auto- and cross-correlator of the current in the different spin channels in first non-zero order of the expansion over the spin-spin interaction.
The current operator for the spin channel $\alpha$ is $I_{\alpha} = \dot{N}_{\alpha}=-i[N_{\alpha},H]_{-}$, where $N_{\alpha}=\sum_{p}c_{\alpha p}^{\dagger}c_{\alpha p}$ is the number of particles in the spin channel $\alpha$.
Introducing the tunneling operators
\begin{equation}
\jmath_{\uparrow} = \frac i2\mathsf{t} \gamma c_{\uparrow}, 
\qquad
\jmath_{\downarrow} = i \mathsf{t} d^{\dagger} c_{\downarrow}, 
\end{equation}
the current operators and Hamiltonian become
\begin{align}
\label{eq:Ilj}
I_{\uparrow/\downarrow} \!=& \,\jmath_{\uparrow/\downarrow}(1-i\lambda\jmath_{\downarrow/\uparrow}^{\dagger}) + h.c. \\
\label{eq:Hlj}
H \!=& \!-\!i(\jmath_{\uparrow}+\jmath_{\downarrow}) \!+\! h.c. \!-\!\lambda \jmath_{\uparrow}\jmath_{\downarrow}^{\dagger} \!-\! \lambda\jmath_{\uparrow}^{\dagger}\jmath_{\downarrow} \!+\!  \frac\lambda2 (\jmath_{\uparrow}\jmath_{\uparrow}^{\dagger}d^{\dagger}d + \jmath_{\uparrow}^{\dagger}\jmath_{\uparrow}dd^{\dagger})
     \!-\! 2 \lambda (\jmath_{\downarrow}\jmath_{\downarrow}^{\dagger} + \jmath_{\downarrow}^{\dagger}\jmath_{\downarrow}) 
\end{align}
where $\lambda = 
2J/\mathsf{t}^{2}$.
The term in the last line of the Hamiltonian corresponds to the $z$-component spin coupling\footnote{%
Note that \mbox{$\jmath_{\downarrow}\jmath_{\downarrow}^{\dagger}d^{\dagger}d=\jmath_{\downarrow}\jmath_{\downarrow}^{\dagger}$}, \mbox{$\jmath_{\downarrow}^{\dagger}\jmath_{\downarrow}dd^{\dagger}=\jmath_{\downarrow}^{\dagger}\jmath_{\downarrow}$}, and \mbox{$\jmath_{\downarrow}\jmath_{\downarrow}^{\dagger}dd^{\dagger}=\jmath_{\downarrow}^{\dagger}\jmath_{\downarrow}d^{\dagger}d=0$}.} %
given in the third line of Eq.~\eqref{eq:H}, and, as we can see, preserves the number of particles in both electron spin channels as well as the occupation of the effective fermion state $d$.
Thus, it cannot contribute to the cross correlations to lowest order perturbation theory in $J$.
Therefore, we will neglect it in the following.

\section{Current-current correlators}

The current fluctuations in classical electronics are considered as a hinderance of the perfect functioning device, but sometimes can be used to study further the internal physics the studied systems~\cite{Heikkilae2013}.
Performing the time-resolved current measurement one can extract various correlators of the current. 
Here we study the generalization of the noise---the current-current correlator defined as
\begin{equation}
S_{\alpha\beta}(t-t') = \langle I_{\alpha}(t) I_{\beta}(t') + I_{\beta}(t') I_{\alpha}(t) \rangle,
\end{equation}
which is simply the Keldysh component of the Green's function $\langle\mathrm{T}_\mathsf{K} I_{\alpha}(t_{i}) I_{\beta}(t'_{j})\rangle$, see Appendix~\ref{apx:contour}.
The angle brackets denote an ensemble average over all the quantum ground state of the system.
This quantum average is generally equivalent to a time average according to the ergodic hypothesis.
For a single spin channel, i.e.\ when $\alpha=\beta$, it is equivalent to the current noise. 
The signal sent through the spectrum analyser allows to obtain the frequency-dependent correlators.
In the case of the noise it can be expressed to zeroth order in $J$ as
\begin{equation}
S_{\alpha\alpha}(\omega) = \frac{e^{2}}{\hbar^{2}}\Bigl( \Pi_{\alpha}^{K}(\omega) + [\Pi_{\alpha}^{K}(-\omega)]^{*} \Bigr),
\label{eq:S1ch}
\end{equation}
where the index $K$ denotes the Keldysh components of the correlators
\begin{equation}
\Pi_{\alpha}(t_{i},t_{j}') = 
\Bigl\langle\mathrm{T}_\mathsf{K}
 \left(\jmath_{\alpha}(t_{i})+\jmath_{\alpha}^{\dagger}(t_{i})\right)\jmath_{\alpha}(t_{j}')
\Bigr\rangle.
\label{eq:Pi1def}
\end{equation}
Here, the correlators are defined on the Keldysh contour~\cite{RammerSmith1986} so that indices $i,j=\pm$ correspond to upper/lower branch, see Appendix~\ref{apx:contour}.
The $\Pi_{\alpha}$ can be expressed by means of normal ($G$ and $G^{+}$) and anomalous ($F$) lead Green's functions, localized state correlator ($D$), and electron cross-correlators ($W$ and $W^{+}$) defined on the Keldysh contour, i.e.
\begin{align}
G_{\alpha}(t_{i},t_{j}') &= -i
\left< \mathrm{T}_\mathsf{K}c_{\alpha}(t_{i})c_{\alpha}^{\dagger}(t_{j}') \right>, &
G_{\alpha}^{+}(t_{i},t_{j}') &= -i
\left< \mathrm{T}_\mathsf{K}c_{\alpha}^{\dagger}(t_{i})c_{\alpha}(t_{j}') \right>, \nonumber\\
F_{\alpha}(t_{i},t_{j}') &= -i
\left< \mathrm{T}_\mathsf{K}c_{\alpha}(t_{i})c_{\alpha}(t_{j}') \right>, &
D_{\alpha}(t_{i},t_{j}') &= -i
\left< \mathrm{T}_\mathsf{K}d_{\alpha}(t_{i})d_{\alpha}^{\dagger}(t_{j}') \right>, \label{eq:Gfdef}\\
W_{\alpha}(t_{i},t_{j}') &= -i
\left< \mathrm{T}_\mathsf{K}c_{\alpha}(t_{i})d_{\alpha}^{\dagger}(t_{j}') \right>, &
W_{\alpha}^{+}(t_{i},t_{j}') &= -i
\left< \mathrm{T}_\mathsf{K}c_{\alpha}^{\dagger}(t_{i})d_{\alpha}(t_{j}') \right>, \nonumber
\end{align}
%
where $d_{\downarrow}=d$ and $d_{\uparrow}=\gamma/2$.\footnote{%
Here we chose the prefactor $1/2$ in order to unify the formulae.
The residue of the Majorana Green's function in this case is $1/2$.
Note that in case of definition $d_{\uparrow}=\gamma$ the residue would be $2$.
}
Here, the ``plus'' symbol in the superscript means $X^{+R/A}(\omega)=-[X^{A/R}(-\omega)]^{*}$ and $X^{+K}(\omega)=[X^{K}(-\omega)]^{*}$ where $X$ can be $G$, $W$ or $\Pi$ (will be defined below).
Bare (in the expansion over the tunneling $\mathsf{t}$) electron Green's functions are expressed as
\begin{equation}
G^{(0)R/A}_{\alpha}(\omega) = \mp i\pi\nu_{\alpha},
\qquad
G^{(0)K}_{\alpha}(\omega) = -2\pi i\nu_{\alpha}\mathrm{th}(\omega),
\end{equation}
where $\nu_{\alpha}$ is the density of states and $\mathrm{th}(\omega)=\tanh\frac{\omega-\mu_{\alpha}}{2T}$.
Full Green's functions of electrons and their cross-correlators where the tunneling $\mathsf{t}$ is taken into account exactly) can be obtained through the full localized state Green's function by means of the Dyson equations
\begin{equation}
\hat{G}_{\alpha}(\omega) = \hat{G}_{\alpha}^{(0)}(\omega)+\mathsf{t}^{2}\hat{G}_{\alpha}^{(0)}(\omega)\hat{D}_{\alpha}(\omega)\hat{G}_{\alpha}^{(0)}(\omega),
\qquad
\hat{W}_{\alpha}(\omega) = \mathsf{t}\hat{G}_{\alpha}^{(0)}(\omega)\hat{D}_{\alpha}(\omega).
\label{eq:GWdyson}
\end{equation}
where the hat denotes the matrix structure in Keldysh space (see Appendix~\ref{apx:contour}).
The full localized state Green's function can be written as
\begin{equation}
D_{\alpha}^{R/A}(\omega)=\left[ [D_{\alpha}^{(0)R/A}(\omega)]^{-1}-\Sigma_{\alpha}^{R/A}(\omega)\right]^{-1}\!\!,
\quad
D_{\alpha}^{K}(\omega)=D_{\alpha}^{R}(\omega)\Sigma_{\alpha}^{K}(\omega)D_{\alpha}^{A}(\omega).
\label{eq:DfullRAK}
\end{equation}
Note that the Keldysh component, i.e.\ the occupation of the localized state is determined by the Keldysh component of the self-energy, which in turn is determined by the distribution function of the channel.
Physically this manifests the fact that the spin channels are massive leads and therefore their distribution function is decisive for the occupation of the localized state. 
Let us calculate the localized state Green's function and correlator $\Pi$ for both spin channels.


For the spin ``down'' channel the Hamiltonian corresponds to the spinless non-interacting resonant-level model, so the bare dot Green's function is
\begin{equation}
D_{\downarrow}^{(0)R/A}(\omega) =\left[\omega\pm i0\right]^{-1}
\label{eq:D0f}
\end{equation}
and the exact self-energy is given by the relation
\begin{equation}
\hat\Sigma_{\downarrow}(\omega) =  \mathsf{t}^{2} \hat{G}_{\downarrow}^{(0)}(\omega).
\end{equation}
Defining the energy scale $\Gamma=\pi\nu\mathsf{t}^{2}$, we get
\begin{equation}
\Sigma_{\downarrow}^{R/A}(\omega) = \mp i \Gamma,
\qquad
\Sigma_{\downarrow}^{K}(\omega) = - 2i\Gamma \mathrm{th}(\omega).
\end{equation}
The correlator $\Pi$ is then equal to
\begin{equation}
\hat\Pi_{\downarrow} = \mathsf{t}^{2}\Bigl(
\hat{D}_{\downarrow}\circ\hat{G}_{\downarrow}- \hat{W}_{\downarrow}\circ\hat{W}_{\downarrow}
\Bigr),
\label{eq:PiFdiag}
\end{equation}
see Appendix~\ref{apx:noise} for details, in particular the definition of the convolution $\circ$.


The spin ``up'' channel is coupled to a single Majorana state, the bare Green's function of which is
\begin{equation}
D_{\uparrow}^{(0)R/A}(\omega) = \left[2\omega\pm i0\right]^{-1}.
\label{eq:D0m}
\end{equation}
The exact self-energy for the Majorana state can be written as
\begin{equation}
\hat{\Sigma}_{\uparrow}(\omega) =  \mathsf{t}^{2}\left( \hat{G}_{\uparrow}^{(0)}(\omega) + \hat{G}_{\uparrow}^{+(0)}(\omega)\right),
\end{equation}
which in terms of the same rate $\Gamma = \pi \nu \mathsf{t}^{2}$ becomes
\begin{equation}
\Sigma_{\uparrow}^{R/\!A\!}(\omega) \!=\! \mp 2i \Gamma, 
\qquad
\Sigma_{\uparrow}^{K\!}(\omega) \!=\! - 2i \Gamma \left(\mathrm{th}(\omega) \!-\! \mathrm{th}(-\omega)\right).
\end{equation}
In addition to the Dyson equations in Eq.~\eqref{eq:GWdyson}, the Majorana fermion creates non-zero anomalous correlations in the spin channel equal to
\begin{equation}
\hat{F}_{\uparrow}(\omega) = -\mathsf{t}^{2}\hat{G}_{\uparrow}^{(0)}(\omega)\hat{D}_{\uparrow}(\omega)\hat{G}_{\uparrow}^{+(0)}(\omega).
\end{equation}
The correlator $\Pi$ thus gets more terms yielding
\begin{equation}
\hat{\Pi}_{\uparrow} = \mathsf{t}^{2}\Bigl(
\hat{D}_{\uparrow}\circ(\hat{G}_{\uparrow}+\hat{F}_{\uparrow}) - (\hat{W}_{\uparrow}+\hat{W}_{\uparrow}^{+})\circ\hat{W}_{\uparrow}
\Bigr),
\label{eq:PiMdiag}
\end{equation}
see Appendix~\ref{apx:noise} for further specifications.


The cross-correlator of the currents in the different spin channels being calculated in first order expansion in $\lambda$ consists of two contributions
\begin{equation}
S_{\uparrow\downarrow}(\omega) = \frac{e^{2}}{\hbar}
\sum_{\genfrac{}{}{0pt}{}{a=1,2}{\alpha\beta=\uparrow\downarrow,\downarrow\uparrow}}\Bigr(
\Pi_{\alpha\beta}^{(a)K}(\omega) + [\Pi_{\alpha\beta}^{(a)K}(-\omega)]^{*}
\Bigl),
\label{eq:Suddef}
\end{equation}
which correspond to the ($a=1$) correction to the expression for the current operator in Eq.~\eqref{eq:Ilj}, and ($a=2$) the interaction terms in the Hamiltonian, see Eq.~\eqref{eq:Hlj}.
Using the diagrammatic technique, we can demonstrate that they are equal to
\begin{align}
\Pi_{\alpha\beta}^{(1)K}(\omega) =  i\lambda\Pi^{K}_{\alpha}(\omega)\langle\jmath_{\beta}^{\dagger}\rangle,
\qquad
\Pi_{\alpha\beta}^{(2)K}(\omega) = -i\lambda\Pi^{K}_{\alpha}(\omega)[\Pi^{R}_{\beta}(\omega)]^{*},
\label{eq:Pidef}
\end{align}
%
%
%
for details see Appendix~\ref{apx:ccorr}.

\section{Results}


Looking back at the Eq.~\eqref{eq:H} we see that in the limit of zero Kondo coupling $J$ the system splits into two independent ones: the spin ``up'' channel is coupled to the Majorana fermion only, while the spin ``down'' channel and effective conventional fermion $d$ form a non-interacting resonant level model (RLM).
All correlators and cross-correlators cancel out at zero frequency even for finite temperatures
Therefore we state the results at zero temperature and zero bias, but for finite frequency $\omega$.
On the basis of Eqs.~\eqref{eq:S1ch}--\eqref{eq:DfullRAK} together with Eqs.~(\ref{eq:D0f}--\ref{eq:PiFdiag}) we can extract the expression for the noise for the RLM
\begin{equation}
S_{\downarrow\downarrow}(\omega) \!=\!
\frac{e^{2}}{h}\frac{8|\omega|\Gamma}{\omega^{2}+4\Gamma^{2}}
\!\left(\!
\omega\arctan\frac{\omega}{\Gamma} \!+\! \Gamma\log \!\left[\!1 \!+\! \frac{\omega^{2}}{\Gamma^{2}}\!\right]
\right).\!\!
\label{eq:SresF}
\end{equation}
Furthermore, with the help of Eqs.~\eqref{eq:D0m}--\eqref{eq:PiMdiag}, we obtain the noise in the channel with the Majorana state coupled to it
\begin{equation}
S_{\uparrow\uparrow}(\omega) = \frac{e^{2}}{h}4\Gamma \arctan\frac{|\omega|}{\Gamma}.
\label{eq:SresM}
\end{equation}

\begin{figure}
\centering
\includegraphics[scale=.8]{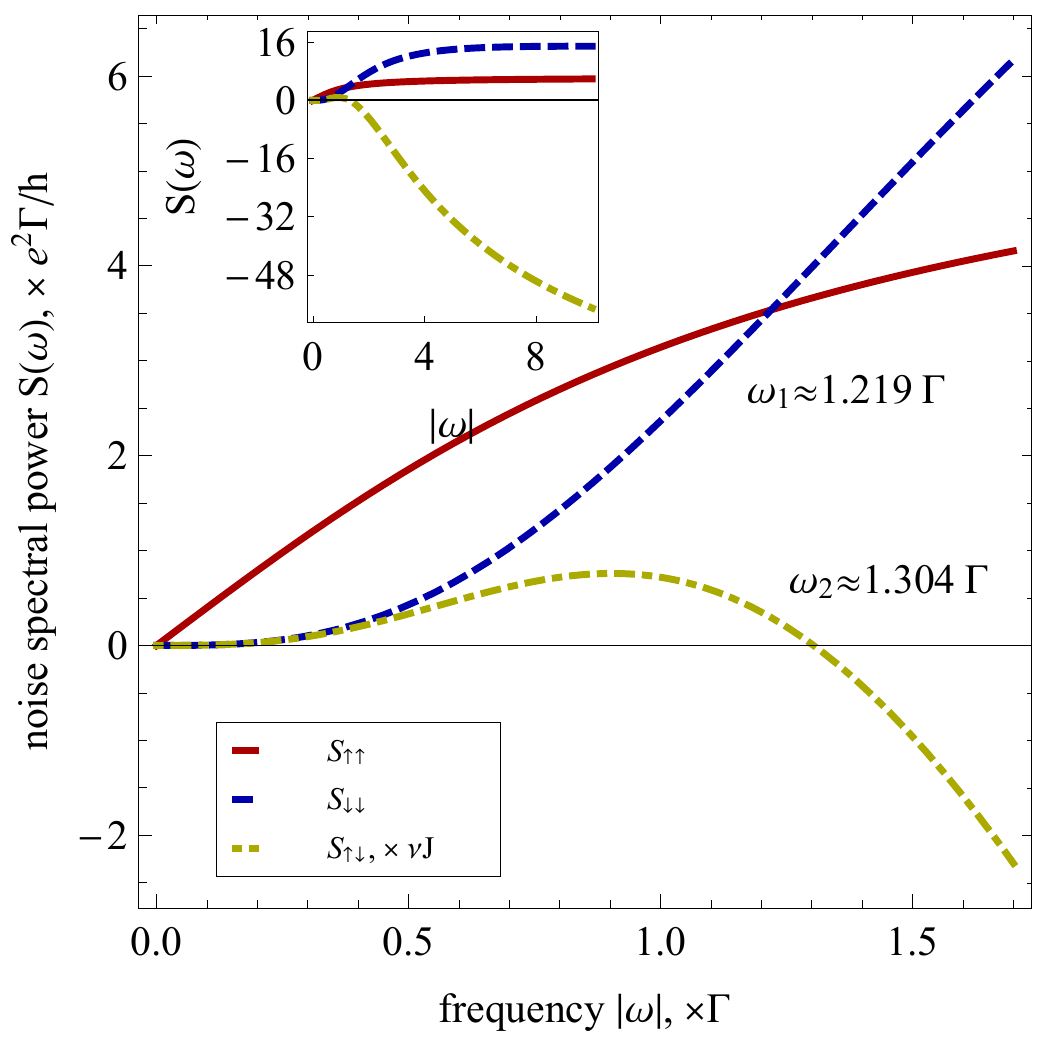}
\caption{The spin currents correlations and their cross-correlator.
The Majorana current corellator (red, solid) dominates at small frequencies, but looses to the RLM noise (blue, dashed) at short times.
The critical frequency $\omega_{1}$ is a universal value proportional to the tunneling coupling $\Gamma$.
The cross-correlations (yellow, dot-dashed) change their sign from positive to negative after another universal frequency $\omega_{2}$.}
\label{fig:correlator}
\end{figure}

The cross-correlator between the currents in the different channels is equal to 
\begin{multline}
S_{\uparrow\downarrow}(\omega)=
\frac{e^{2}}{h} \nu J \frac{4|\omega|\Gamma}{\omega^{2}+4\Gamma^{2}}\times\\\times
\Biggl(
4\Gamma \arctan^{2}\frac{\omega}{\Gamma}
-
2\omega \arctan\frac{\omega}{\Gamma}\log\left[1+\frac{\omega^{2}}{\Gamma^{2}}\right]
-
\Gamma \log^{2}\left[1+\frac{\omega^{2}}{\Gamma^{2}}\right]
\Biggr).
\label{eq:SresK}
\end{multline}
The behavior of all three current correlators is represented in the Fig.~\ref{fig:correlator}.

The expansion of the expressions in Eqs.~\eqref{eq:SresF}--\eqref{eq:SresK} at small $\omega\ll\Gamma$ frequencies demonstrate the same cubic dependence for the non-interacting resonant-level model and the cross-correlator.
The Majorana coupled to the lead, however, causes stronger, linear in $\omega$ noise.
These results can be summarized as
\begin{equation}
S_{\alpha\beta} \approx \frac{4e^{2}}{\hbar}|\omega| 
\begin{cases}
 \omega^{2}/\Gamma^{2} & \alpha\beta=\downarrow\downarrow,
\\
\nu J \omega^{2}/\Gamma^{2} & \alpha\beta=\uparrow\downarrow,
\\
 1 & \alpha\beta=\uparrow\uparrow.
\end{cases}
\end{equation}
At large frequencies $\omega\gg\Gamma$, the noise in the resonant-level model, is, contrarily, twice larger than in the Majorana case.
While the strength of both noises gets saturated for large frequencies, the cross-correlator grows logarithmically:
\begin{equation}
S_{\alpha\beta} \approx \frac{e^{2}}{\hbar}\Gamma
\begin{cases}
2 +(8\Gamma/\pi\omega)\log(\omega/\Gamma) & \alpha\beta=\downarrow\downarrow,
\\
1 -4\Gamma/\omega & \alpha\beta=\uparrow\uparrow,
\\
 - 4\nu J\log(\omega/\Gamma) & \alpha\beta=\uparrow\downarrow.
\end{cases}
\end{equation}

Let us discuss the physical interpretations of the obtained results.
The result for the spin-down channel in the zero order in Kondo coupling corresponds to the spinless lead coupled to the single state quantum dot, which is described by the resonant-level model (RLM).
Therefore the current correlation in the spin-down channel is equal to the noise in the RLM setup.
At short times the correlator of the spin-up channel, which is coupled to the Majorana fermion, is twice small than the correlator for the spin-up channel coupled to the conventional fermion.
This fact looks natural, since two Majoranas constitute a single conventional fermion, so can be seen as a ``half'' of it.
At large times the correlations naturally decay, but differently for the conventional and Majorana fermion.
Majorana state cannot be blocked according to the Pauli principle as a conventional fermion state, the current noise in the spin-up channel, which is coupled to the Majorana, is stronger than in the spin-down channel, which is coupled to the single fermion state.

The sign of the cross-correlations is different for the times much larger than the typical time $\Gamma^{-1}$ that reflects the role the relaxation processes.
%
The cross-correlator can be represented as $S_{\uparrow\downarrow} = \lambda S_{\uparrow\uparrow} R_{\downarrow\downarrow} + \lambda S_{\downarrow\downarrow} R_{\uparrow\uparrow}$, where in time representation $R= \mathrm{Re} [i \Pi^{R}(t)]$ is a current response of the RLM excited from equilibrium by transfer of a single electron from the channel to the localized state at time $t=0$.
At large times, the contribution $\lambda S_{\uparrow\uparrow} R_{\downarrow\downarrow}$ dominates and demonstrates the same frequency dependence as the resonant-level model.
The cross-noise can be seen as a response of the RLM to the stronger noise created by the Majorana state.
Since RLM has weaker correlations, it is the ``bottleneck'' in this process, and therefore the cross-correlations inherit the frequency dependence from the RLM correlations.

\section{Summary}

We have calculated the auto- and the cross-correlations of the currents in the different spin channels of a helical edge state coupled to the localized topological Kondo impurity.
We have also demonstrated that the cross-correlator of the currents in different spin channels changes its sign depending on the time scale of the correlations, what may indicate the change of the statistics of the collective excitations.
The comparison of the noise at large time scales demonstrated the same frequency dependence as in the resonant-level model.
The transport measurements in the proposed experimental setup, namely the signals collected from the different sides of the quantum point contact (marked ``input'' and ``output'' in the Fig.~\ref{fig:setup}), allow the direct measurement of the frequency resolved noise and cross-correlations.
The sign change of the cross-correlation at $\omega_{2}$ and the crossing of the noise levels at $\omega_{1}$ can be used as a proof of the Majorana bound state presence.
The measuring of the values of these frequencies ($\omega_{1/2}$) gives the value of the tunneling coupling $\Gamma$.
The value of the cross-correlations at large frequencies allows the extraction of the Kondo coupling $J$.
Thus, the results obtained in this work allow to interpret the noise measurements in the experimental setups similar to Ref.~\cite{Deng2016} justifying the presence of the Majorana fermion and its connection to the rest of the system.

%

\begin{acknowledgement}
Financial support by the DFG grant SFB1170 "ToCoTronics".
Author thanks B.~Trauzettel for the stimulating discussions.
\end{acknowledgement}

\appendix


\section{Keldysh contour}
\label{apx:contour}

We follow here the notations of Rammer and Smith review~\cite{RammerSmith1986}, where the Keldysh contour consists out of two parts: ($c_{1}\equiv+$) line lies \emph{above} the time axis and is the \emph{first} part of the Keldysh contour, while ($c_{2}\equiv-$) line lies \emph{below} the time axis and is the \emph{second} part of the contour.
Thus the Green's function of two operators $A$ and $B$ defined as 
\begin{equation}
G_\mathsf{K}(t_{i},t_{j}') = -i\langle\mathrm{T}_\mathsf{K}A(t_{i})B(t_{j}')\rangle,
\end{equation}
where $\mathrm{T}_\mathsf{K}$ is the time ordering along the Keldysh contour and $i=\pm$ denotes the upper/lower part.
It places the ``$t_{-}$'' operator always on the left, independent on the value of $t$ and $t'$ if the latter one sits on the ``+'' branch, i.e.\ $G_\mathsf{K}(t_{-},t_{+}')=G^{>}(t,t')=-i\langle A(t)B(t')\rangle$.
The four elements of the Green's function corresponding to the different signs of $i$ and $j$ are linearly dependent.
Thus, one should make the transformation in the Keldysh space in order to obtain the retarded (R), advanced (A), and Keldysh (K) components of the Green's function, i.e.
\begin{equation}
\hat{G}_{ij}(t-t') = \sum_{i'j'}(L\tau^{3})_{ii'}G_\mathsf{K}(t_{i},t_{j}')(L^\mathsf{T})_{j'j},
\quad
L=\frac{1-i\tau^{2}}{\sqrt2},
\label{eq:Krotation}
\quad
\hat{G}=
\begin{pmatrix}
G^{R} & G^{K} \\
0 & G^{A}
\end{pmatrix},
\end{equation}
where $\tau^{1/2/3}$ are Pauli matrices.

%

\section{Noise in single channel}
\label{apx:noise}

\begin{figure}
\centering
\includegraphics[page=1,scale=.35]{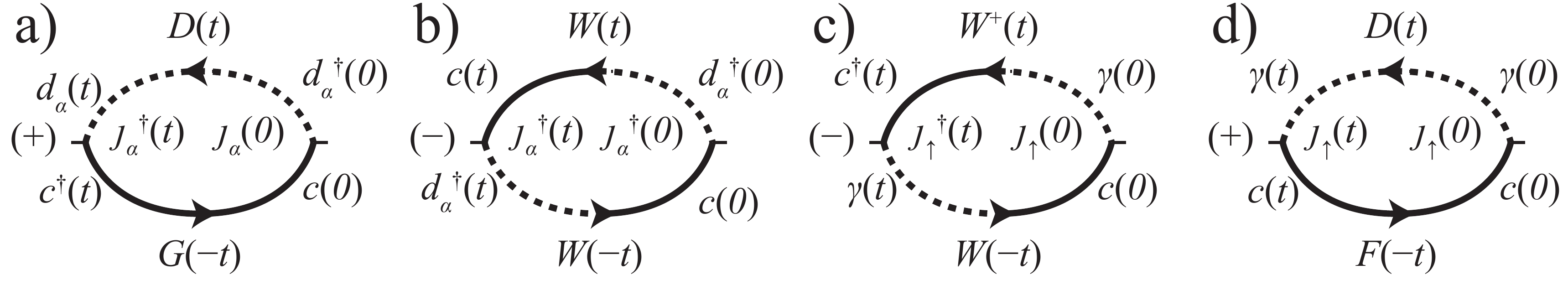}
\caption{
Diagrams for correlators $\Pi_{\downarrow}$ and $\Pi_{\uparrow}$.
a,b) Diagrams for correlators $\Pi_{\alpha}$ corresponding to the terms in Eq.~\eqref{eq:PiFdiag}.
The operator $d_{\alpha}$ can be identified with either $d$ (for $\alpha=\downarrow$) or $\gamma/2$ (for $\alpha=\uparrow$).
c,d) Additional diagrams for correlator $\Pi_{\uparrow}$ corresponding to the terms in Eq.~\eqref{eq:PiMdiag} that are missing in Eq.~\eqref{eq:PiFdiag}.
The Green functions appearing in this diagram are defined in Eqs.~\eqref{eq:Gfdef}, and sign in brackets denotes the sign before the term obtained according to the Wick theorem.}
\label{fig:diagramMF}
\end{figure}

The noise for a single spin channel can be explicitly calculated by Eq.~\eqref{eq:Pi1def} and Wick theorem.
For the spin-``down'' channel the only non-zero terms are demonstrated in the Fig.~\ref{fig:diagramMF}a,b).
In the Keldysh contour representation, the arguments of both Green's function coincide up to second order in tunneling.
For Fig.~\ref{fig:diagramMF}a) the diagram contributing to $\Pi_\mathsf{K}(t_{i},0_{j})$ equals, for instance, $\mathsf{t}^{2} D_\mathsf{K}(t_{i},0_{j})G_\mathsf{K}(0_{j},t_{i})$.
Performing the rotation in Keldysh space described in Eq.~\eqref{eq:Krotation}, the components get mixed, so in the new representation the contribution to $\hat{\Pi}(t)$ is
\begin{equation}
\mathsf{t}^{2} \bigl(\hat{D}\circ\hat{G}\bigr)(t) = \bigl(\hat{D}(t)\circ\hat{G}(-t)\bigr)
\end{equation}
where the convolution sign means
\begin{align}
2\bigl(\hat{D}\circ\hat{G}\bigr)^{R/A} &= D^{R/A}G^{K}+D^{K}G^{A/R},
\\
2\bigl(\hat{D}\circ\hat{G}\bigr)^{K\hphantom{/A}} &= D^{R}G^{A}+D^{A}G^{R}+D^{K}G^{K}.
\end{align}
Changing to the frequency representation, the contribution to $\hat{\Pi}(\omega)$ can be written as
\begin{equation}
\mathsf{t}^{2}\bigl(\hat{D}\circ\hat{G}\bigr)(\omega) = \mathsf{t}^{2}\int\bigl(\hat{D}(\omega'+\omega)\circ\hat{G}(\omega')\bigr)\frac{d\omega'}{2\pi}.
\end{equation}
The second diagram in Fig.~\ref{fig:diagramMF} is calculated in the same way.

For the spin-``up'' channel there are two more non-zero terms, which are illustrated in the Fig.~\ref{fig:diagramMF}c,d).
They originate from the anomalous tunneling in Fig.~\ref{fig:diagramMF}c) and the anomalous correlation in Fig.~\ref{fig:diagramMF}d).

\section{Cross-correlator}
\label{apx:ccorr}

\begin{figure}
\centering
\includegraphics[page=2,scale=.35]{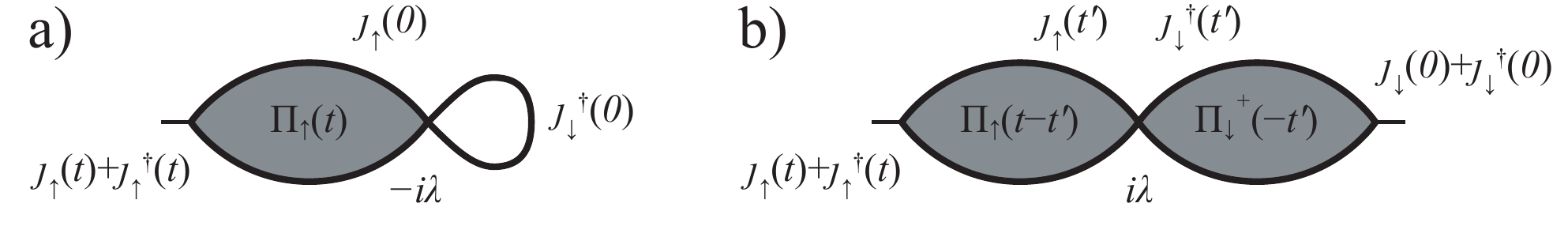}
\caption{
The diagrams for the cross-correlator:
the diagram for a) contribution $\Pi_{\uparrow\downarrow}^{(1)}$ and b) contribution $\Pi_{\uparrow\downarrow}^{(2)}$, see Eq.~\eqref{eq:Pidef}.}
\label{fig:diagramK}
\end{figure}

The cross-correlator terms linear in $\lambda$, have two different origins, which come from (I) correction to the expression for the current operator in Eq.~\eqref{eq:Ilj}, and (II) the interaction terms in the Hamiltonian, Eq.~\eqref{eq:Hlj}.
Taking the current operator correction $-i\lambda\jmath_{\downarrow}\jmath_{\uparrow}^{\dagger}$ out of the expression for $I_{\downarrow}$, we get the diagram shown in the Fig.~\ref{fig:diagramK}a), which we denote as $\Pi^{(1)}_{\uparrow\downarrow}(\omega)$.
The gray loop is the single channel correlator $\Pi_{\uparrow}$ defined in Eq.~\eqref{eq:Pi1def}, the other loop is just the average of the operator $\jmath_{\downarrow}^{\dagger}$, and the crossing thick lines denote the vertex $\lambda$, so the net expression is
\begin{equation}
i\lambda\hat{\Pi}_{\uparrow}(t) \langle \jmath_{\downarrow}^{\dagger} \rangle.
\end{equation}
The other contribution $i\lambda\jmath_{\uparrow}\jmath_{\downarrow}^{\dagger}$ results in the diagram $[\Pi^{(1)}_{\uparrow\downarrow}(-\omega)]^{*}$.
The same corrections coming from the operator $I_{\uparrow}$ will simply correspond to the $\Pi^{(1)}_{\downarrow\uparrow}(\omega)$ and $[\Pi^{(1)}_{\downarrow\uparrow}(-\omega)]^{*}$

The second type of the correction comes from the interaction terms $-\lambda \jmath_{\uparrow}\jmath_{\downarrow}^{\dagger}$ and $- \lambda\jmath_{\uparrow}^{\dagger}\jmath_{\downarrow}$ in the Hamiltonian, Eq.~\eqref{eq:Hlj}.
The rest of the interaction terms do not contribute in the first order of the expansion in $\lambda$.
Taking the term $-\lambda \jmath_{\uparrow}\jmath_{\downarrow}^{\dagger}$, we get the diagram shown in Fig.~\ref{fig:diagramK}b) which is equal to
\begin{equation}
-i\lambda\int \hat{\Pi}_{\uparrow}(t-t') \tau^{1}\hat{\Pi}^{+}_{\uparrow}(-t')\tau^{1} dt',
\end{equation}
where the ``plus'' superscript denotes again $\Pi^{+R/A}(\omega)=-[\Pi^{A/R}(-\omega)]^{*}$ and $\Pi^{+K}(\omega)=[\Pi^{K}(-\omega)]^{*}$.
The matrix $\tau^{1}$ takes into account the fact that the definition of $\Pi^{+}$ is $\langle\mathrm{T}_\mathsf{K}
 (\jmath_{\downarrow}(t_{i})+\jmath_{\downarrow}^{\dagger}(t_{i}))\jmath_{\downarrow}^{\dagger}(t_{j}')\rangle$, while in the diagram the operators are swapped.
The frequency representation removes the integration over time.
Then, the Keldysh component of the diagram is
\begin{equation}
-i\lambda \Pi^{K}_{\uparrow}(\omega) \Pi^{+A}_{\uparrow}(-\omega) -i\lambda \Pi^{R}_{\uparrow}(\omega) \Pi^{+K}_{\uparrow}(-\omega).
\end{equation}
Together with the other interaction term $-\lambda\jmath_{\uparrow}^{\dagger}\jmath_{\downarrow}$ we get the full set of  diagrams relevant to evaluate Eqs.~\eqref{eq:Suddef} and~\eqref{eq:Pidef}.


\bibliographystyle{epj}
\bibliography{tknoise}

\end{document}